\documentstyle[prl,aps,multicol,epsfig]{revtex}

\begin{document}
\title{Resistivity saturation revisited: results from a dynamical mean field theory}
\author{A. J. Millis}
\address{Department of Physics and Astronomy, John Hopkins University,\\Baltimore, MD 21218}
\author{Jun Hu and S. Das Sarma}
\address{Department of Physics, University of Maryland, College\\
Park, MD 20742}
\maketitle
\begin{abstract}
We use the dynamical mean field method to study the high-temperature
resistivity of electrons strongly coupled to phonons. The results reproduce
the qualitative behavior of the
temperature and disorder dependence of the resistivity of  'A-15' materials 
which is
normally described in terms of saturation, but imply that
the resistivity does not saturate.  Rather, a change in temperature
dependence occurs when the scattering becomes strong enough
to cause a breakdown of the Migdal approximation. {\it PACS: 72.10-d,72.10.Di,72.80.Ga}

\end{abstract}

\begin{multicols}{2}

The conventional \cite{Bloch} theory of 
the temperature dependent resistivity of metals
predicts that at temperatures greater than the phonon
Debye frequency $\omega_D$  the resistivity, $\rho $,
behaves as 
\begin{equation}
\rho =A_{\rho }T+B_{\rho }  \label{Bloch}
\end{equation}
with the temperature coefficient A$_{\rho }$ 
proportional to the electron-phonon coupling strength and 
residual resistivity B$_{\rho }$ proportional
to the impurity concentration. The theory
accounts well for the properties of many metals but fails in a number of
important cases. The first  to attract widespread
attention were the 'A-15' materials such as $V_{3}Si$
and $Nb_{3}Ge$.  In these compuinds 
the electron-phonon coupling is unusually strong and although the 
high temperature  $\rho (T)$ may be crudely fit to Eq. 1, 
$A_{\rho}$ is much smaller than expected and, crucially,  $B_{\rho}$ is
large even in pure samples \cite{Fisk76}.  
Further, the effects of extra disorder (induced
e.g. by radiation damage) are not additive: the extra
disorder-induced resisitivity is temperature dependent, being larger at
lower T and smaller at higher T; indeed for sufficiently high
disorder levels $\rho(T)$ becomes flat and then
$d\rho/dT$ becomes negative \cite{Lutz76}.  This 
complex of behaviors was named 'resistivity
saturation' by Fisk and Webb \cite{Fisk76}.
Interest in this long-standing problem 
has been increased by
observations that
in a number of
materials in which electron-electron
interactions are believed to be important, including 
the high-T$_{c}$ cuprates\cite{Gurvitch86}, 
the  $A_{3}C_{60}$ materials \cite{Palstra94}, and
$Sr$RuO$_{3}$\cite{Klein96},  $\rho (T)$ increases rapidly
with increasing temperature and exhibits no signs of saturation even in the 
sense defined above, although the
resistivity becomes much larger than that of A-15s.  
A recent paper \cite{Emery95} argued that this 
{\it absence of saturation} was itself anomalous and interesting.

Ref \cite{Fisk76} has led to a large but
still inconclusive literature on the cause of the phenomenon\cite
{Allen,Belitz,Gurvitch,Yu}.
Eq \ref{Bloch} is the mathematical statement that
$\rho$ is proportional to lattice disorder, i.e. that  
electrons are scattered by 
displacements, due either to thermal fluctuations or defects in the crystal,
of ions (labelled by $i$) from their ideal crystallographic positions.  
Its derivation
is based on three assumptions: (1) at
temperatures greater than $\omega_D$, the thermal
fluctuations of the ions may be modelled in terms of  classical harmonic
oscillators with spring constant $k$;
equipartition then implies $\left\langle x_{i}^{2}\right\rangle \sim T/k$
plus a T-independent term arising from defects, (2) 
the electron-ion interaction
may be treated by second-order perturbation theory so that if the 
electron-ion coupling is $g$ the
electron scattering rate is $g^{2}\left\langle x_{i}^{2}\right\rangle $ and (3)
usual Boltzmann transport theory may be used to relate the resistivity
to the scattering rate. Combining these considerations yields Eq \ref{Bloch}
with $A\sim g^{2}/k$ and B proportional to the impurity concentration.

Deviations from Eq \ref{Bloch} must come from the breakdown of one of the
assumptions.  Many authors have focussed on (3), however
attempts to generate a
systematic series of 'multiple scattering'
corrections to the Boltzmann equation
have produced a multiplicity of terms whose
interpretation and range of validity have 
not been clear (see, e.g. \cite{Allen}). However, 
examination of assumption (3) 
led Allen and Chakraborty to an important observation
concerning the {\it magnitude} of the resistivity\cite{Allen}. In the A-15
materials, the carriers are d-electrons and although the total
carrier density  is
4 d-electrons per metal ion, i.e. $n=12$ per formula unit,
the band structure is such that
only one d-band crosses the fermi surface implying $n=1$ per formula unit.
The remaining d electrons are
contained in filled bands which are
very close to the fermi energy and to empty bands. 
A moderately large scattering rate will mix the full and empty bands 
with the conduction band, leading
to a drastic increase in the effective number of carriers. 
Ref \cite{Allen}
concluded that while the effect does not 
fully explain the observed 'saturation'
behavior, it does permit the
magnitude of the high temperature resistivity to vary dramatically
between different systems.  We note here that the ruthenates and 
cuprates do not have  many filled and empty d-bands near the fermi surface,
and so may have a lower carrier
density and thus higher resistivity than do the A-15s.  

Belitz and Schirmacher \cite{Belitz} used a 
'memory function' formalism to study a model of 
electrons interacting with phonons
via both density and stress couplings.  In their model $\rho$ increased 
indefinitely
with T due to an interplay between
scattering caused by the density coupling and hopping 
induced by the stress coupling. Our results 
suggest that the stress coupling and associated hopping
are not necessary to reproduce
the phenomenon.

Yu and Anderson\cite{Yu} focussed instead on
assumptions (1) and (2), arguing that a
strong electron-phonon coupling could lead to a spontaneously generated
double-well potential for the phonon, rendering
the second-order perturbation theory treatment invalid and
requiring a sophisticated quantum treatment.
Our results support the idea that a strong electron-phonon coupling
is the essence of the problem but imply that neither
double-well formation nor quantum effects are essential.

In this Letter we 
calculate the resistivity of a model of electrons coupled with arbitrary 
strength to phonons. 
Because we seek to understand the fundamental mechanism underlying the 
saturation phenomenon we consider the simplest 
possible model: spinless electrons coupled to
dispersionless phonons and to static disorder.
Because saturation is a high-T phenomenon
we further specialize to  classical phonons ($\omega_D \rightarrow 0$).
The Hamiltonian $%
H=H_{el}+H_{ph}+H_{el-ph}+H_{disorder}$ with electronic part $%
H_{el}=-\sum_{p}\epsilon_pd_{p}^{+}d_{p}$ describing
motion of electrons in a band with dispersion $\epsilon_p$
phonon part $H_{ph}=\sum_{i}\frac{1}{2}kx_{i}^{2}$, 
an electron-phonon coupling 
$H_{el-ph}=g\sum_{i}x_{i}(d_{i}^{+}d_{i}-n)$, 
and an disorder part 
$H_{disorder}=\sum_{i}w_{i}(d_{i}^{+}d_{i}-n).$ 
The form for $H_{disorder}$ corresponds
to electrons scattered by random point defects which may be thought of 
as frozen-in lattice distortions of amplitude $w_i/g$.  The
mean density of electrons is $n$ so $x=0$ is the equilibrium phonon state
for a uniform distribution of electrons. We assume the random site 
energies $w_{i}$ associated with the disorder

are distributed
according to $P_{dis}(w)=\exp (-w^{2}/\eta ^{2})/(\sqrt{2\pi \eta ^{2}}).$

We define the parameter $t$ to be one quarter of the 
full electron bandwidth;
we measure all energies and temperatures in units of $t$.  
Rescaling the phonon coordinate $x \rightarrow x/\sqrt{k}$ 
shows that the dimensionless
parameter describing the electron-phonon coupling 
is $\lambda = g^2/(kt)$ (this 
$\lambda$ is a factor of $\pi$ larger than the 
conventional MacMillan parameter).
At low T the model may be solved by the usual Migdal pertubation theory,
which is an expansion in 
$\lambda \sqrt{\max (\omega_{D},T}/t$ \cite{Migdal}.
To treat arbitrary coupling strengths we employ
the dynamical mean field approximation \cite{Georges96} which  
becomes {\it exact} in the limit of spatial dimensionality 
$d\rightarrow \infty $ and which comparisons to other techniques and to
experiment have shown to be quantitatively accurate in $d=3$. 
The approximation may be derived from the assumption 
that the electron
Green function $G$ is 
\begin{equation}
G(\epsilon_p,i\omega _{n})=\left[ i\omega _{n}-\Sigma (i\omega _{n})-\epsilon
_{p}+\mu \right] ^{-1}.  \label{eq6}
\end{equation}
where  the self-energy
$\Sigma $
is taken to be $p$-independent.  This ansatz implies
that all interaction effects are derivable from
the local ($p$-integrated) Green function, $G_{loc}$ given by 
$G_{loc}(i\omega _{n})=\int \frac{d^{d}p}{(2\pi )^{d}}\;G(\epsilon_p,i\omega
_{n})=\int \;\frac{d\epsilon _{p}{\cal D}(\epsilon _{p})}{i\omega
_{n}-\Sigma (i\omega _{n})-\mu -\epsilon _{p}}.$ 
G$_{loc}$ is itself given in terms of a mean field function
$a(\omega)$ determined by an equation which depends on
the density of states ${\cal D}$; we choose
the semicircular form ${\cal D}_{semi}(\epsilon _{p})=\sqrt{4t^{2}-\epsilon
_{p}^{2}}/(2\pi \;t^{2}),$ so that
\begin{equation}
a(\omega )=\omega +\mu -\int_{-\infty }^{\infty }dxdw\frac{P_{phonon}(x,w)P_{dis}(w)}{%
a(\omega )+gx+w}  \label{a}
\end{equation}
with the phonon probability distribution 
\begin{equation}
P_{phonon}=\frac{1}{Z_{loc}}exp\left[ -\frac{x^{2}}{2T}+\int d\omega 
\;ln\;\left[ a(\omega)+gx+w)\right] \right]  \label{eq16}
\end{equation}
with $Z_{loc}=\int dxdw P_{phonon}(x,w)P_{dis}(w)$.
$G_{loc}(\omega)=\frac{\partial \;ln\;Z_{loc}}{\partial a(\omega)}$
and $\Sigma(\omega)=a(\omega)-G_{loc}^{-1}(\omega)$.  This treatment
of static disorder is equivalent to the familiar coherent potential

We solve the equations numerically;
computational details are given in ref\cite{Millis96a}.
The conductivity may be calculated from \cite{Georges96}

\begin{equation}
\sigma \simeq \int \;d\omega \;\int d\epsilon _{p}{\cal D}(\epsilon
_{p})[Im\;G(\epsilon _{p},\omega ]^{2}\;cosh^{-2}\frac{\omega }{2T}
\label{eq18}
\end{equation}
(Ref \cite{Georges96} takes the current operator to
be independent of $\varepsilon _{p}$; this does not affect results in any
important way).

\vspace{0.25cm}
\centerline{\epsfxsize=8cm \epsfbox{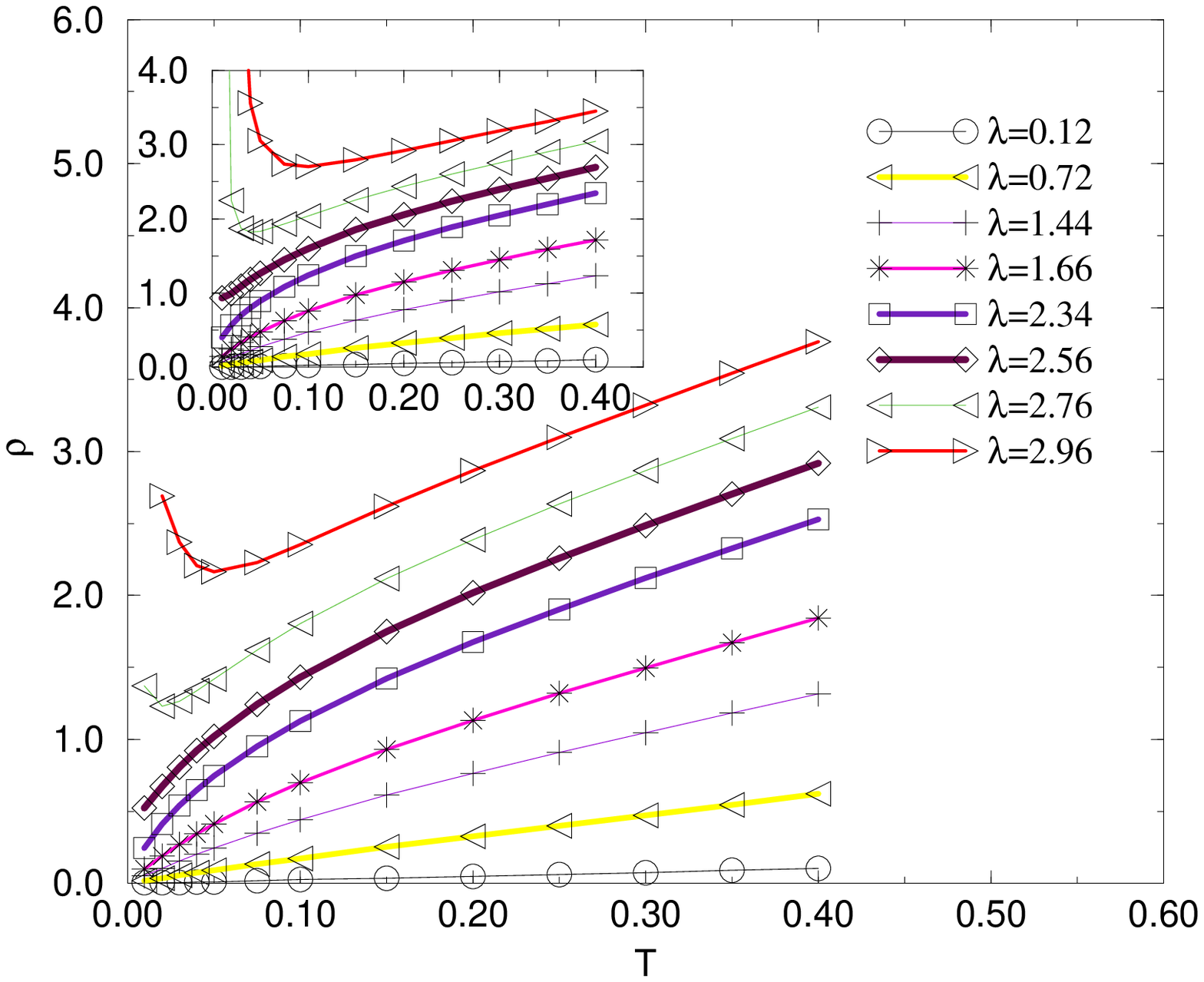}}
{\footnotesize {\bf Fig 1} Resistivity $\rho (T) $ vs temperature
T (in units of t=bandwidth/4) at various 
couplings $\lambda$ for density n=1/4.
Inset: Same couplings, density n=1/2.}
\vspace{0.25cm}

Fig. 1 shows the calculated resistivity $\rho $ vs. temperature $T$ at various
couplings (assuming $\eta =0$ i.e. no disorder) for density n=1/4; the inset shows the
same calculation for n=1/2. Clearly the two fillings display the same qualitative
behavior: at weak coupling, the resistivity
has the usual linear T-dependence with a zero offset and a
slope proportional to $\lambda $.  (Quantum effects
neglected here would cause the resistivity to drop 
dramatically once T is
reduced below $\omega _{D}$ but do not 
affect\cite{Migdal} the behavior at $T>\omega
_{D}$; our results are only meaningful for $T>\omega_D$). 
For the narrow bands typical of A-15 materials, $T=0.1$
corresponds to a temparture of order 300K, and only the results
for $T > 0.1$ should be regarded as physically meaningful.
At  intermediate coupling, the high-temperature resistivity displays the 
essential features of the 'saturation' behavior found in the data, namely
an apparently linear T dependence with non-zero offset and
slope rather weakly dependent on $\lambda$.   At very strong coupling, 
the high-T behavior is
not much changed (except that the offset becomes larger relative to the
slope) but at low T the system becomes insulating ($d\rho/dT<0$) below
a $\lambda$- dependent characteristic temperature $T_{gap}$
because a gap of size $T_{gap}$ opens
at low T in the electron spectral
function.  
Our interest is in the high-T behavior; the low-T
insulating regime is extensively discussed in Ref \cite{Millis96a}.

We now determine the origin of the saturation behavior. The first issue is
the relation between the resistivity and the electron self-energy. Fig 2
plots the calculated resistivity versus the imaginary part of the electron
self-energy evaluated at the 
fermi surface, ($\Sigma ^{\prime \prime}(\omega =0))$. 

\vspace{0.25cm}
\centerline{\epsfxsize=8cm \epsfbox{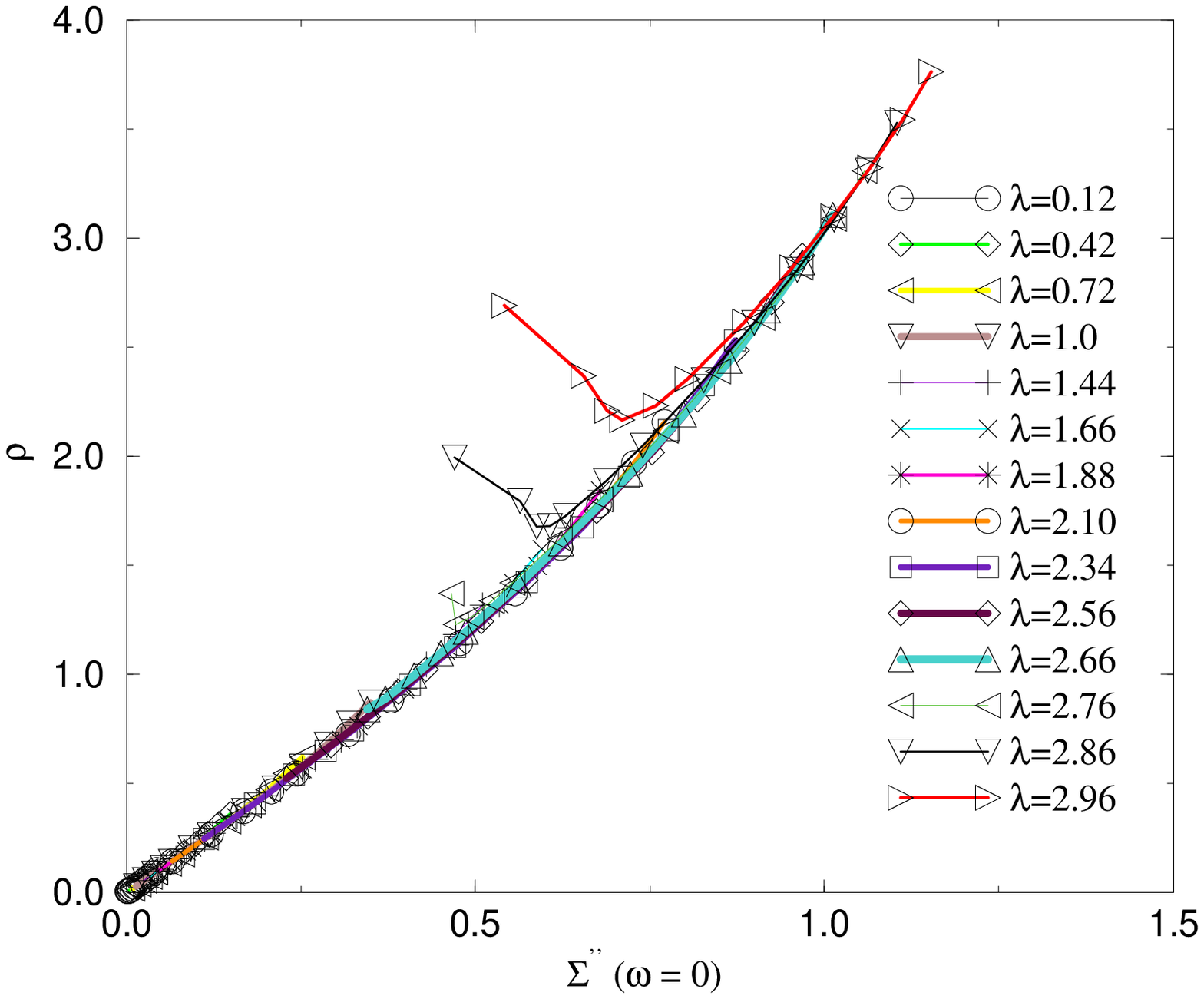}}
{\footnotesize {\bf Fig 2} Resistivity $\rho $ plotted against fermi surface scattering
rate $\Sigma^{^{\prime\prime}}$ (given in units of t=bandwidth/4)
for various couplings, with temperature
as an implicit parameter.  The points which break away from 
the universal curve correspond
to the low T insulating regime.}
\vspace{0.25cm}

One sees that except in the insulating
low-T strong coupling gapped regime, $\rho $ is a {\it universal} 
function of $\Sigma^{\prime \prime}(0) $. At
weak coupling, $\rho \sim \Sigma^{\prime \prime}(0) $ 
as expected; at stronger couplings $\rho $
increases {\it faster} than $\Sigma^{\prime \prime}(0)$. Thus assumption (3)
is not the issue.

We next consider phonon anharmonicity. Fig 3 shows the mean square
displacement of the oscillator coordinate as a function of temperature for
the coupling strengths used in Fig. 1.At weak coupling one sees the
classical behavior $\left\langle x^{2}\right\rangle \sim T$ with zero
intercept. As the coupling is increased the high-T behavior acquires the
form $\left\langle x^{2}\right\rangle \sim A_{phonon}T+B_{phonon}$. However,
a comparison of Figs 1 and 3 shows the ratio B$_{phonon}/A_{phonon}$ 
is much less than the ratio B$_{\rho }/A_{\rho }$; in
other words, the resistivity curves are much flatter than the lattice
displacement curves. 
\vspace{0.25cm}
\centerline{\epsfxsize=8cm \epsfbox{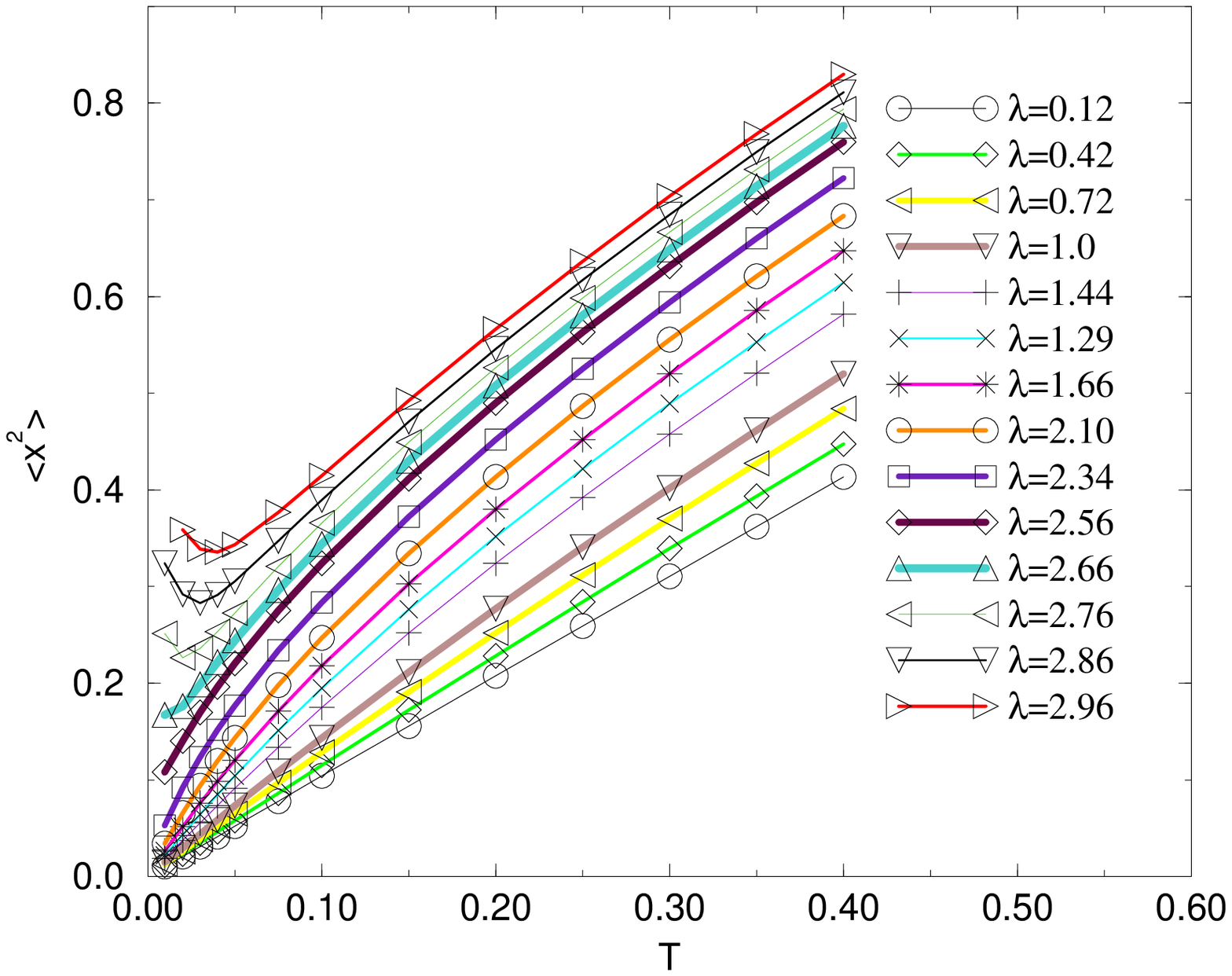}}
{\footnotesize {\bf Fig 3} Mean square lattice 
displacement $<x^2>$ vs. temperature $T$ 
in units of t=bandwidth/4 at various
couplings.}
\vspace{0.25cm}

Further, the calculated phonon probability distribution
$P(x)$ (not shown) reveals that the
double-well form discussed by Yu and Anderson occurs only in the low T,
large g 'insulating' regime.  Thus 
anharmonicity, while quantitatively important, is not the 
fundamental cause
of the phenomenon.

The remaining issue is assumption (2): the relation of the self-energy to the
scattering mechanism. The  weak-coupling (Migdal) result for
our model is $\Sigma ^{^{\prime \prime }}(\omega =0)=
g^{2}<x^{2}>(1-(\mu /2t)^{2})^{1/2}$.
To see how this relationship evolves we plot
in Fig 4 $\Sigma ^{^{\prime \prime }}(\omega =0)/g^{2}<x^{2}>$ as a function
of temperature at different couplings. 

\vspace{0.25cm}
\centerline{\epsfxsize=8cm \epsfbox{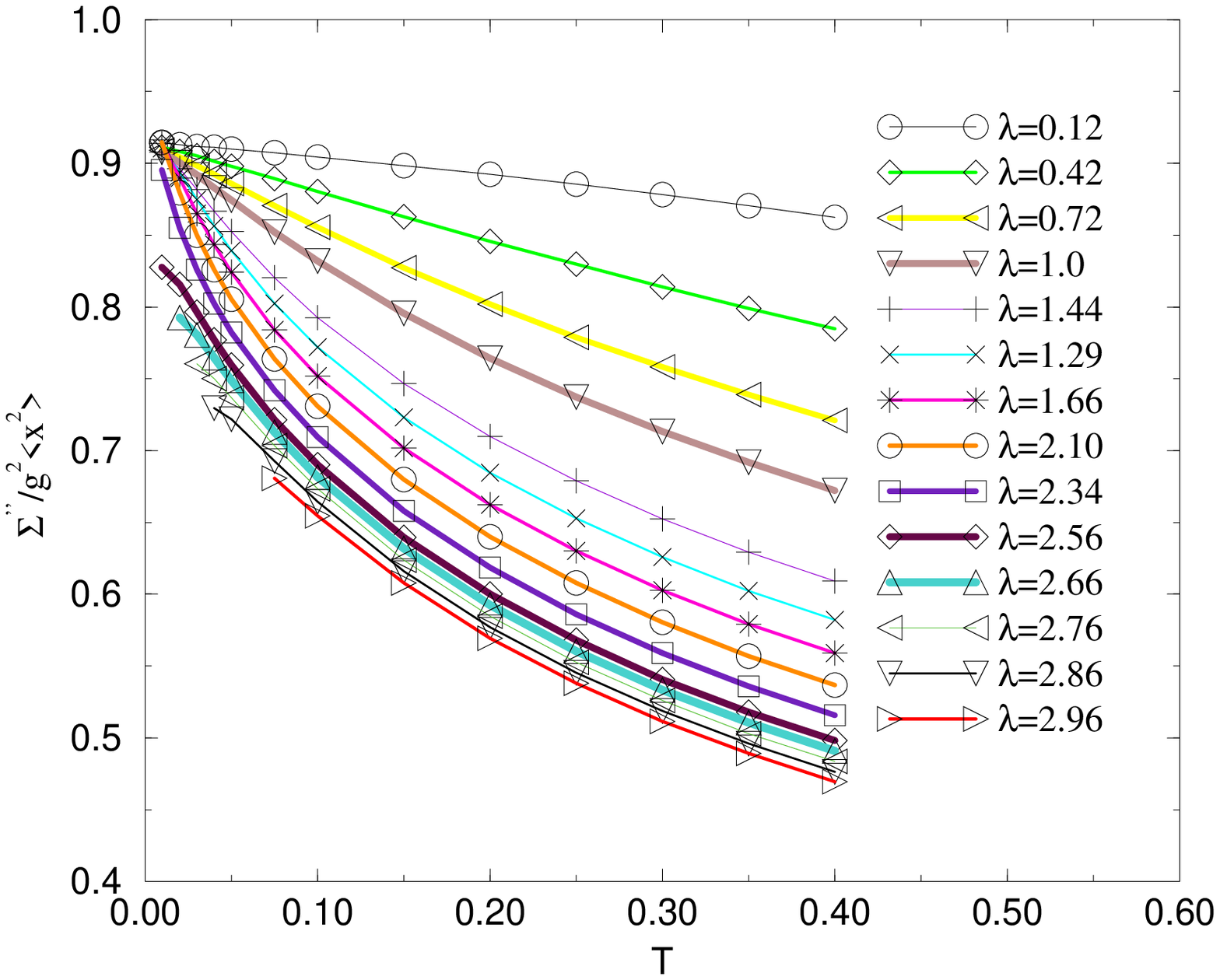}}
{\footnotesize {\bf Fig 4} $\Sigma^{^{\prime\prime}} 
(\omega = 0)/g^2<x^2>$ vs. temperature $T$ at
various couplings.}
\vspace{0.25cm}

We see that as $g^{2}<x^{2}>$
increases, the ratio falls sharply below the weak coupling value and indeed
ultimately $\Sigma ^{^{\prime \prime }}(\omega =0)$ becomes proportional 
to ($g^{2}<x^{2}>)^{1/2}$ i.e. roughly to  $T^{1/2}$.
The crossover may be understood analytically from Eq 
\ref{a}. At low T and small g,$a \sim it$ so one may expand in gx/a, 
generating the
familiar Migdal series. At high T and large g 
a is negligible and the integral is dominated by a pole at $x\sim it/g$;
taking account of the normalization of P(x) yields $G_{loc}\sim
1/g\left\langle x^{2}\right\rangle ^{1/2}$ and so 
$\Sigma \sim g\left\langle
x^{2}\right\rangle ^{1/2}$. {\it Thus the key to saturation
is that at strong coupling the
scattering rate continues to increase
but at a rate less rapid than that given by second order
perturbation theory.} A similar result for the self
energy of a model of carriers coupled to spin fluctuations
was presented in \cite{Chubukov96}.

We now add disorder scattering. Qualitatively one may think of
impurities as adding an extra T-independent term to $\left\langle
x^{2}\right\rangle $, so $\left\langle x^{2}\right\rangle \rightarrow
\left\langle x_{phonon}^{2}(T)+x_{impurity}^{2}\right\rangle $. At low T and
small $\lambda$ the rates add, leading to  Matthiessen's rule 
$\rho=\rho_{phonon}+\rho_{impurity}$.  At high T, physical quantities depend on $%
\left\langle x^{2}\right\rangle ^{1/2}$ so the relative correction due to
impurities is of order $x_{impurity}^{2}/x_{phonon}^{2}(T)$: Matthiessen's
rule does not apply. Fig 5 shows the results of a direct calculation
confirming this idea.  A more sophisticated (non-CPA)
treatment is required to reproduce the observed
negative $d\rho/dT$ at very strong disorder.

\vspace{0.25cm}
\centerline{\epsfxsize=7cm \epsfbox{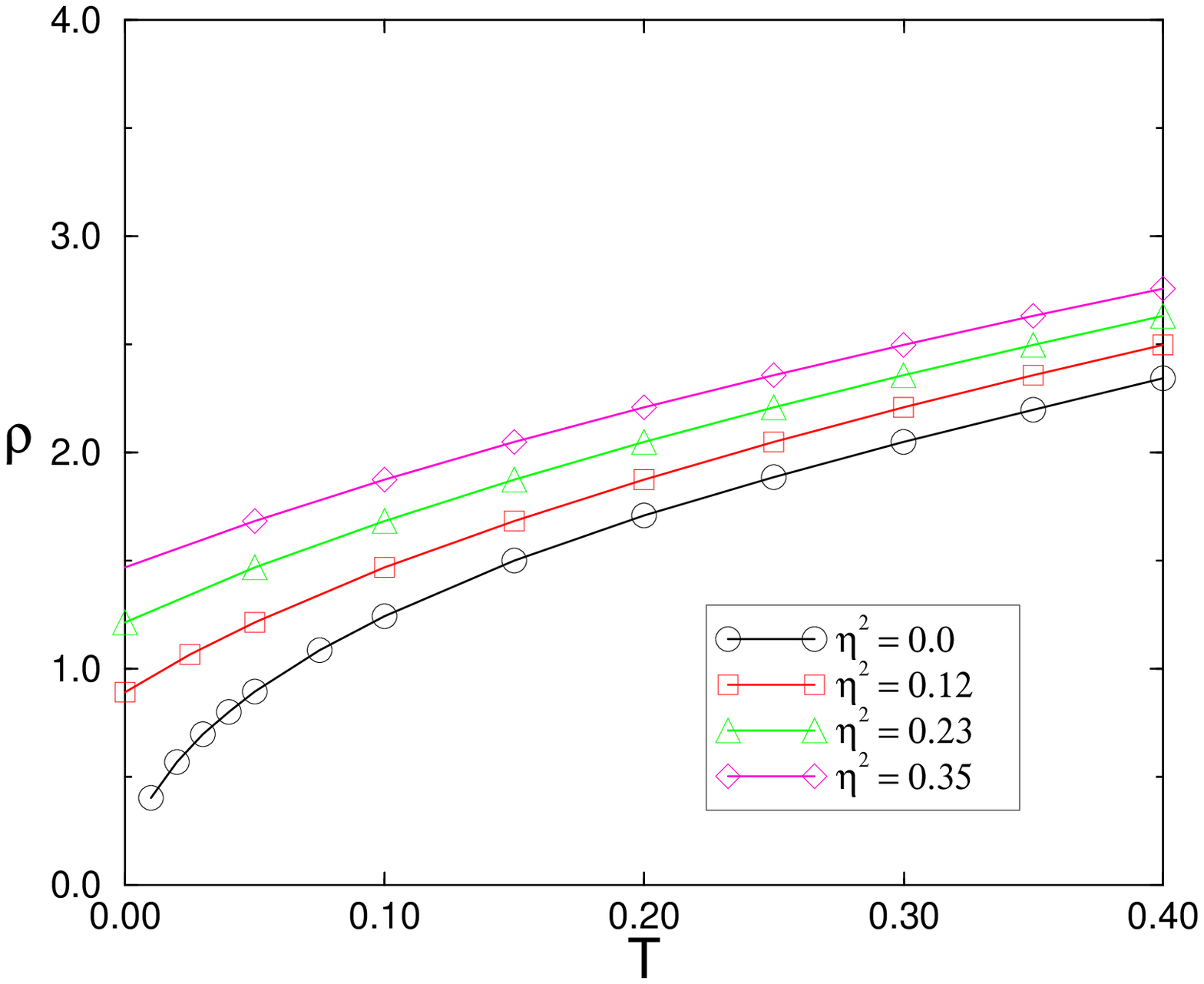}}

\centerline{{\footnotesize {\bf Fig 5} Impurity dependence of resistivity; 
$\lambda=2.34$, n=1/2.}}

\vspace{0.25cm}

To summarize, we have shown that a simple model of electrons coupled to
classical phonons and to static disorder reproduces very well
the essential features
of the phenomenon usually described as resistivity saturation. 
The cause
of the phenomenon was 
found to be a breakdown of the Migdal relation
between lattice distortion amplitude and electron self energy.
Comparison of our results for different carrier
densities along with the band-mixing arguments of Ref \cite{Allen}, 
indicates that the
value of A/B will depend strongly on model details,
so detailed comparison of our calculation to experiments on
specific materials is inappropriate, but material-specific calculations
using the ideas and formalism put forward here would be of great interest.

Our results imply that the term 'saturation' is a misnomer: there is no
intrinsic maximum value of the high-temperature resistivity.
Indeed at very high temperatures 
one expects classical diffusion with  diffusion
constant D vanishing or tending to a constant as $T \rightarrow \infty$, 
implying $\rho$ increasing indefinitely with $T$. This was found
in 'retracable path approximation' studies of the Hubbard model
\cite{Rice88} and recently in 
a model of carriers coupled to magnetically correlated spins
\cite{Parcollet98}. From this perspective the 'absence of saturation' 
\cite{Emery95} in correlated electron materials 
is not by itself surprising; the
interesting issue is the apparent smoothness of the crossover from the
low-T coherent transport regime to the high-T classical diffusion
regime. 

{\it Acknowledgements:} This work was supported by 
NSF-DMR-9707701 (AJM) and
the University of Maryland NSF-MRSEC and the 
US-ARO  (J. H. and S. D-S). 
We thank P. B. Allen, P. W. Anderson,
D. Khmelnitskii, H. Monien and Z. Fisk
for  helpful discussions.  Part
of this work was performed while
A.J.M. visited the Aspen Center for Physics and 
Los Alamos National Laboratory.

\end{multicols}
\end{document}